\documentclass[prd,preprint,tightenlines,floatfix,showpacs,preprintnumbers,nofootinbib,eqsecnum]{revtex4}
 \usepackage[dvips,final]{graphicx}
  \usepackage{amssymb}
   \usepackage{amsmath}
    \usepackage{epsfig}
     \usepackage{bm}

\newcommand{\di}{\displaystyle}
\newcommand{\ga}{\gamma}
\newcommand{\scA}{{\scriptscriptstyle A}}
\newcommand{\si}{\sigma}

\begin{document}
\thispagestyle{empty} \preprint{\hbox{}} \vspace*{-10mm}

\title{High energy Gamma-Ray Bursts as a result of the collapse
and total annihilation of neutralino clumps}

\author{R.~S.~Pasechnik}%
\email{rpasech@theor.jinr.ru}%
\affiliation{Bogoliubov Laboratory of Theoretical Physics, JINR,
141980 Dubna, Russia} \affiliation{Faculty of Physics, Moscow
State University, 119992 Moscow, Russia}

\author{V.~A.~Beylin}
\author{V.~I.~Kuksa}
\author{G.~M.~Vereshkov}
\email{gveresh@ip.rsu.ru}

\affiliation{ Institute of Physics, Rostov State University,
344090 Rostov-on-Don, Russia}

\date{\today}
\begin{abstract}
Rare astrophysical events -- cosmological gamma-ray bursts with
energies over GeV -- are considered as an origin of information
about some SUSY parameters. The model of generation of the
powerful gamma-ray bursts is proposed. According to this model the
gamma-ray burst represents as a result of the collapse and the
total annihilation of the neutralino clump. About 80 \% of the
clump mass radiates during $\sim 100$ second at the final stage of
annihilation. The annihilation spectrum and its characteristic
energies are calculated in the framework of Split Higgsino model.
\end{abstract}

\pacs{}


\maketitle

\section{Introduction}
Since the direct experimental information about structure of SUSY
model and values of SUSY particles' masses are absent now we try
to extract any possible data from the cosmology. From the recent
WMAP data \cite{WMAP} the mass density in the Universe is equal to
its critical value $\rho_{crit}\simeq 0.54\cdot
10^{-5}\,\mathrm{GeV/cm^3}$; the structured Dark Matter (DM) and
the isotropic and homogeneous Dark Energy (DE) contribute to this
density $\simeq 23\%$ and $\simeq 73\%$, correspondingly. As it is
known, baryons give $\approx 4\%$ of the total Universe's mass. In
our Galaxy $\approx 50 \% $ of total mass is the mass of DM which
forms the Halo. The Lightest Supersymmetric Particles (LSP) --
neutralino -- are supposed to be the main contributors to the cold
(and hot) DM. The physics of LSP is discussed intensively in
various aspects \cite{Olive-Nojiri}.

To observe the neutralino appearance in space or land experiments
one can use of the different ways. For example, the high energy
antiprotons, positrons or neutrino flux due to neutralino
annihilation in the Dark Halo or in the center of the Sun or the
Earth. In our work we have concentrated on the specific
gamma-radiation \cite{Cesar} which can be diffuse (due to
annihilation into fermions and bosons) or monochromatic (due to
annihilation into $\gamma \gamma$ or $Z\gamma$ \cite{Ullio}). To
measure characteristics of these specific gamma-fluxes the special
experimental programs have been elaborated with space telescopes
(EGRET, GLAST) \cite{Cesar},~\cite{Feng}.

Now we have from the collider experiments only low restriction for
the neutralino mass $m_{\chi}> 32.5\, \mathrm{GeV}$. ATLAS and CMS
collaborations (LHC) will say something else only after 2007. But
the absolute stability of neutralino (or metastability) cannot be
checked in collider experiments. Moreover, some details of the
standard cosmological scenario, which are based on the neutralino
as the CDM, should be changed if neutralino's and other
superparticles' characteristics do not agree with the scenario.
Thus we need in combined analysis of the LHC and astrophysical
data to verify our ideas on cosmological and cosmogonic evolution
\cite{Ellis1}.

Neutralino in our Galaxy or in other galaxies up to $z=8 \div 10$
at the gravity separation stage can form massive dense clumps of
different sizes \cite{clumps},~\cite{small-scale},~\cite{abund}.
Due to the inner gravitational instability clumps can collapse and
intensively annihilate. Our basic supposition is that the nature
and origin of the rare high-energy GRB can be understood as a
result of the neutralino clump total annihilation. We have shown
that at the final stage of the process the powerful gamma-ray
burst (GRB) with photon energies over GeV will occur. In the
standard GRB the energies of photons are only fractions of MeV
(see \cite{DarRuj}). The specific energy and time distribution of
this hard $\gamma$-radiation can be related with the parameters of
superparticles (neutralino, in particular) and give an important
information on the SUSY model details and its compatibility with
cosmology. As an interesting example we have considered the Split
Higgsino scenario \cite{we} in the framework of famous Split
Supersymmetry model \cite{SS} and obtained the values of
annihilation time of the neutralino clump, characteristic photon
energies of the GRB and the annihilation spectrum.

\section{The collapse of the homogeneous neutralino clump}

Consideration of neutralino as a carrier of DM mass in the
Universe generates questions about their evolution and about
astrophysical phenomenons which prove both the existence of
neutralino and the reality of the evolution process with
neutralino participation.

According to high resolution cold dark matter simulations in the
framework of the hierarchical principle of structure formation the
large virilized halos are formed through the constant merging of
smaller halos produced at earlier times. Neutralino at the
gravitational isolation stage $z=8 \div 10$ formed dense massive
clumps of different scales \cite{small-scale},~\cite{CDM}. It
likes naturally that DM clumps are in the state of dynamical
equilibrium as a result of compensation of own gravitational
forces by forces of tidal interactions with each other. In
consequence of inner instability the clump can turn into
irreversible collapse state. The growth of density provides the
growth of annihilation rate. So neutralino clouds become more and
more intense sources of relativistic particles. At final stage of
this process there is a gamma-ray burst. Specific energy and time
distribution of the gamma radiation are connected with neutralino
parameters and give an important information about structure of
the MSSM and about its correspondence with cosmology. Therefore
our main statement consist in the nature of rare powerful
cosmological gamma ray bursts is a result of the total
annihilation of neutralino clumps.

We have analyzed the qualitative and quantitative descriptions of
this process in the spherically symmetric collapse model with two
assumptions. First, annihilation products leave a clump during a
time substantially smaller than the time of its macroscopic
evolution. And second, an annihilating clump is spatially
homogeneous and isotropic since more dense regions annihilate
faster than the rest and heterogeneities are vanishing.

The model, defined by equations of mass and momentum balance and
the equation of non-relativistic gravitation theory, is:
\begin{eqnarray}
 \begin{array}{c}
  \di \frac{d}{dt}\int \rho_{\chi} dV = -\int \rho_{\chi} v_i dS_i -
  \int \frac{2(\si v)_{ann} \rho_{\chi}^2}{M_{\chi}}dV\,,
\\[5mm]
  \di \frac{d}{dt}\int \rho_{\chi} v_i dV = -\int (\rho_{\chi} v_i
  v_k + P_{ik})dS_k - \int \rho_{\chi} \nabla_i \phi dV - \int
  \nu_{dis}\rho_{\chi} v_i dV - \int \frac{2(\si v)_{ann}
  \rho_{\chi}^2}{M_{\chi}}v_idV\,,
\\[5mm]
  \di \Delta \phi = 4\pi G\rho_{\chi}\,,
\end{array}
\label{e1}
\end{eqnarray}
where $P_{ik}$ is the pressure tensor for neutralino gas. It is
absent in consequence of the second assumption. Last terms in the
balance equations describe the momentary going away of
annihilation products from the collapsing clump. We have taken
into account that annihilation and scattering are cross-channels.
In the local form we have the simple system:
\begin{eqnarray}
  \di \frac{dM}{dt}= -\frac{3(\si v)_{ann}}{2\pi
  M_{\chi}}\cdot \frac{M^2}{R^3}, \qquad
  \left(\frac{dR}{dt}\right)^2 =\frac{2GM}{R}\left(1-\beta^2
  \frac{R}{R_0}\right),
\label{e5}
\end{eqnarray}
which integrates elementary. Here $R$ is the clump radius; $M=4\pi
\rho_{\chi}R^3/3$ is the clump mass; $R_0=const$ is the initial
radius; $\beta^2 \leq 1$ is the constant which fixed by initial
velocity of the surface clump compression. The results are in the
parametric form
\begin{eqnarray}
  \begin{array}{c}
  \di M(\zeta)= M_{(*)}\left[1+\frac{(\si
  v)_{ann}}{M_{\chi}}\left(\frac{3\rho_{\chi(*)}}{2\pi
  G}\right)^{1/2}\left(\frac{\zeta^3}{3}+\beta^2
  \zeta\right)\right]^{-2},
\\[6mm]
  \di R(\zeta)=R_0(\zeta^2+\beta^2)^{-1},
\\[6mm]
  \di \left(\frac{8\pi
  G\rho_{\chi(*)}}{3}\right)^{1/2}t(\zeta)=\frac{1}{\beta^2}\left(\frac{\zeta}{\zeta^2+\beta^2}
  -(1-\beta^2)^{1/2}\right)
  +\frac{1}{\beta^3}\left(\arctan\frac{\zeta}{\beta}
  -\arctan\frac{(1-\beta^2)^{1/2}}{\beta}\right)+
\\[6mm]
  \di +\frac{(\si
  v)_{ann}}{3M_{\chi}}\left(\frac{3\rho_{\chi(*)}}{2\pi
  G}\right)^{1/2}\left(\ln(\zeta^2+\beta^2)
  +\frac{2\beta^2(\zeta^2+\beta^2-1)}{\zeta^2+\beta^2}\right),
\end{array}
\label{e6}
\end{eqnarray}
where $M_{(*)}\sim M_\odot =4\cdot 10^{33}
\mathrm{g},\;\rho_{\chi(*)}\sim 10^{-4} \mathrm{g/cm^3}$ are the
initial clump mass and density. As we see $M(t)\to 0,\; R(t)\to 0$
for $t \to \infty$.

\begin{figure}[!h]
 \begin{minipage}{0.45\textwidth}
 \epsfxsize=\textwidth\epsfbox{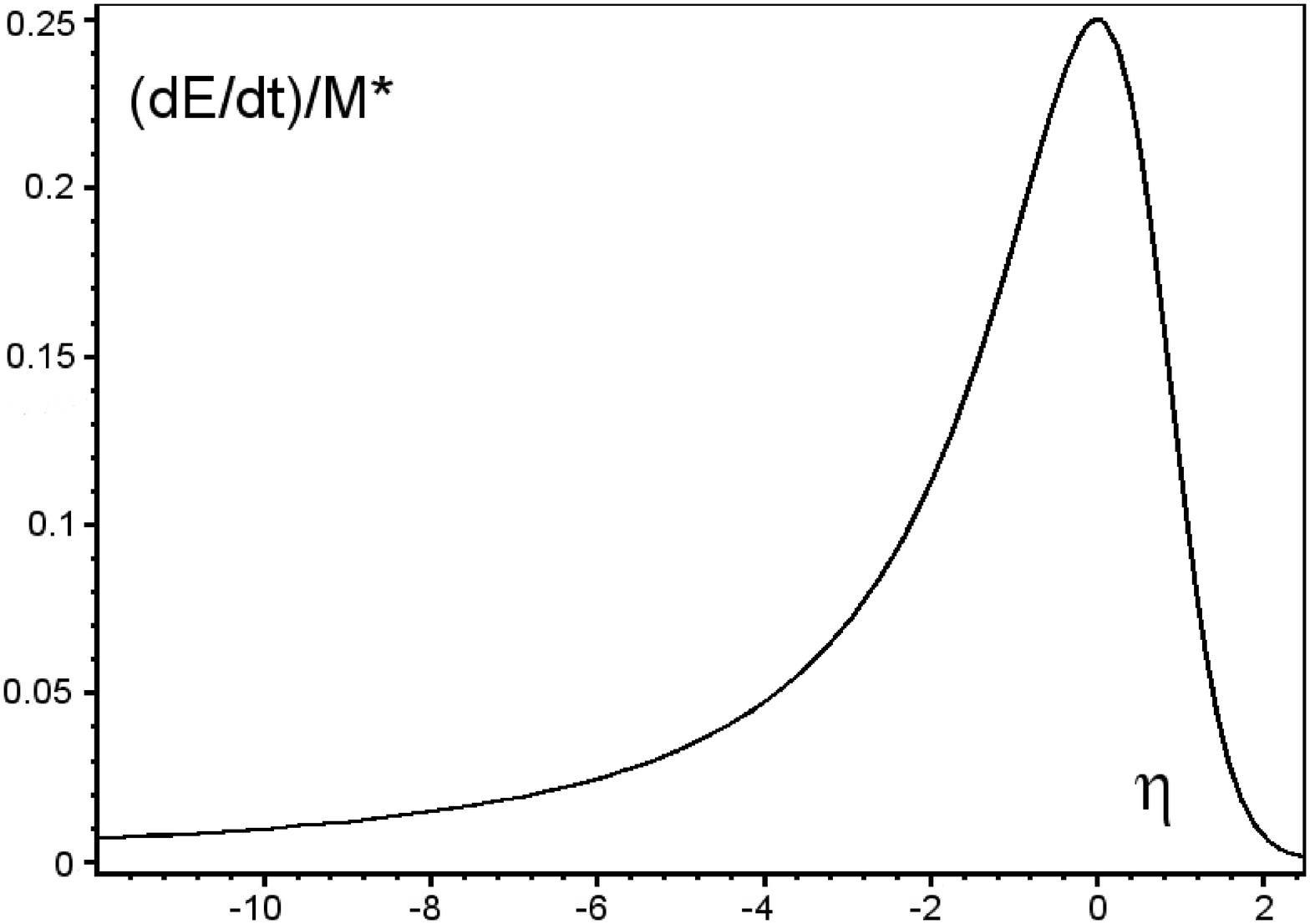}
 \end{minipage}
 \begin{minipage}{0.45\textwidth}
 \epsfxsize=\textwidth \epsfbox{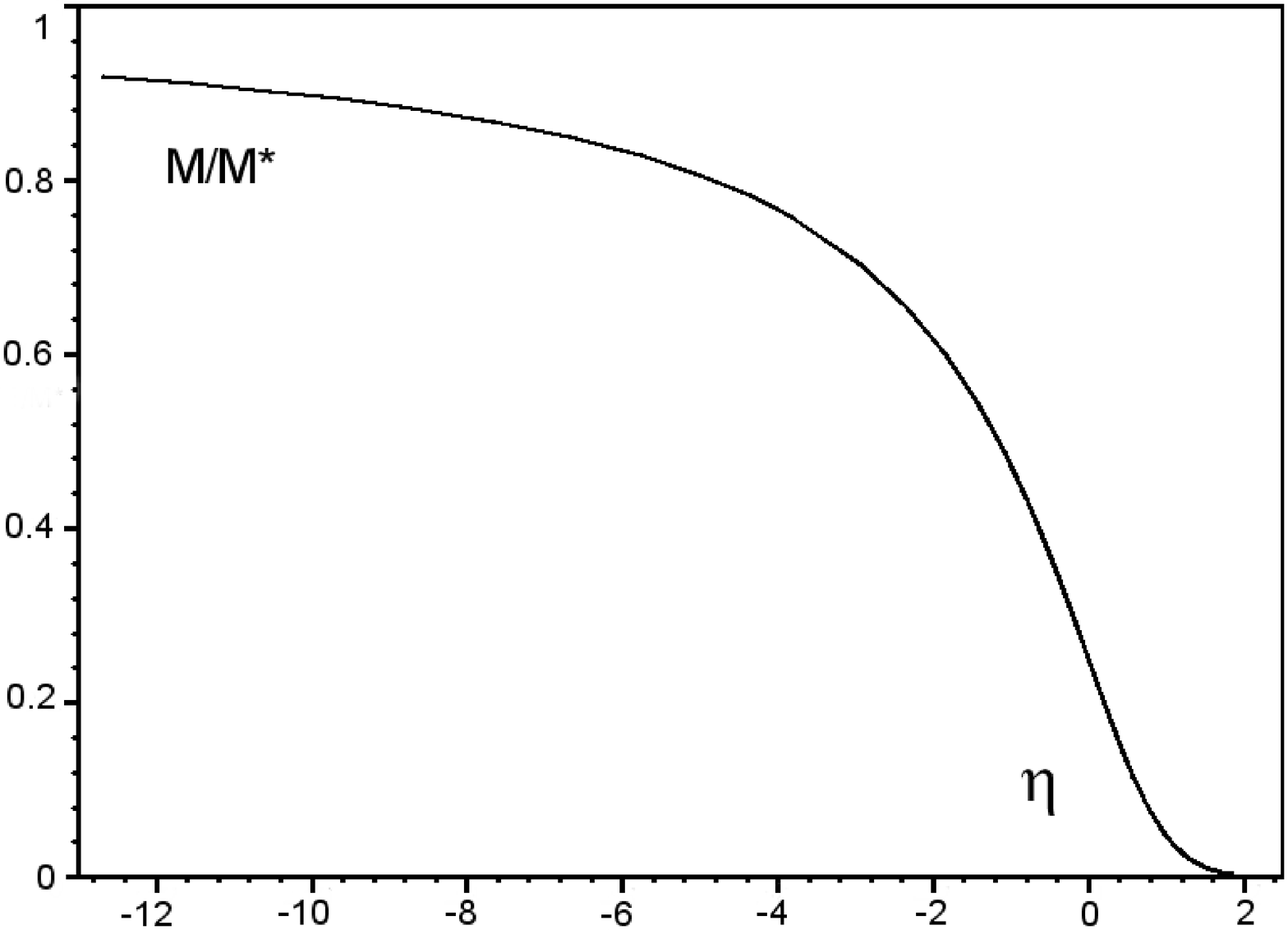}
 \end{minipage}
\caption{\small Annihilation rate $\dot E_A/M_{(*)}$ (right) and
relative mass $M/M_{(*)}$ (left) of neutralino clump as a function
of dimensionless time $\eta$.} \label{fig:fig1}
\end{figure}

The annihilation rate is
\begin{eqnarray}
 \di -\frac{dM}{dt}\equiv
  \dot{E}_{ann} =M_{(*)}\rho_{\chi(*)}\cdot \frac{2(\si
  v)_{ann}}{M_{\chi}} \cdot \frac{(\zeta^2+\beta^2)^3}{\di
  \left[1+\frac{(\si
  v)_{ann}}{M_{\chi}}\left(\frac{3\rho_{\chi(*)}}{2\pi
  G}\right)^{1/2}\left(\frac{\zeta^3}{3}+\beta^2
  \zeta\right)\right]^{4}}.
\label{rad}
\end{eqnarray}
Generally behavior of this function is that a slow growth at small
times changes into quick evolution around narrow peak after that
there is an exponential drop to zero. This behavior explains by
changing of the parameter hierarchy during the collapse. We use
the convenient time scale
\begin{eqnarray}
  \di
  \dot{E}_{ann}=\frac{M_{(*)}}{4}\left(1-\frac{(t-t_0)^2}{2\tau_0^2}\right),
  \qquad \tau_0= \frac{(\si v)_{ann}}{3\pi G M_{\chi}}. \label{max}
\end{eqnarray}
Here $t_0$ is the time of the maximum. Since we have
\[
  \di g=\frac{(\si
  v)_{ann}}{M_{\chi}}\left(\frac{3\rho_{\chi(*)}}{2\pi
  G}\right)^{1/2}\ll 1
\]
then the annihilation rate doesn't depend on initial conditions
and expresses through the fundamental constants only. The
functions $\dot{E}_{ann}/M_{(*)}$ and $M/M_{(*)}$ of dimensionless
time $\eta = (t-t_0)/\tau_0$ are shown at the Fig.~\ref{fig:fig1}.

About 80 \% of clump mass annihilates during the time
\begin{eqnarray}
 \di t_{ann}=6\tau_0=
 \frac{2(\si v)_{ann}}{\pi G
 M_{\chi}}.
 \label{tau}
\end{eqnarray}
Effectively the annihilation takes place between
$t_{\scA}^{(-)}=4\tau_0$ (before maximum) and
$t_{\scA}^{(+)}=2\tau_0$ (after maximum). Such event represents as
a powerful gamma-ray burst with total energy about clump mass.
Kinetic cross section can be parameterized in the form: $(\sigma v
)_{ann}=\alpha^2/8\pi M_{c}^2$, $M_{c}$ -- some parameter with the
dimension of mass. So the decay time of collapsing neutralino
clump expresses through SUSY parameters $M_{c},\,M_{\chi}$ and
gravitational constant but it is absolutely independent from
initial density and initial velocity of clump compressing.

\begin{figure}[h]
 \centerline{\includegraphics[width=0.95\textwidth]{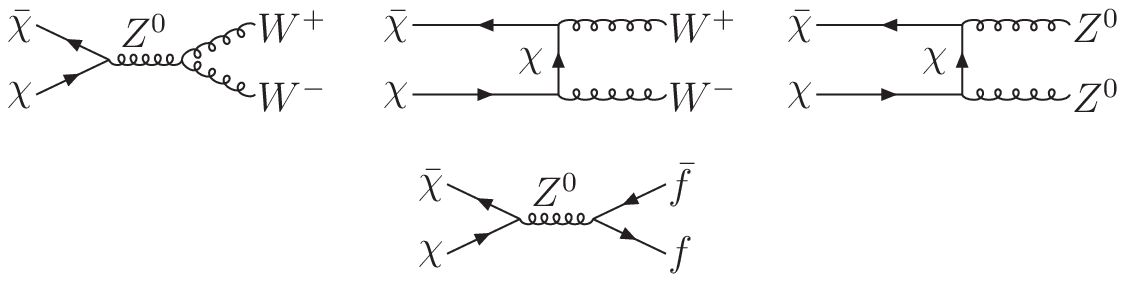}}
   \caption{\label{fig:fig2}
   \small Diagrams of neutralino annihilation in the
   framework of Split Higgsino scenario.}
\end{figure}

Deeper analysis is possible only in the framework of concrete SUSY
model. So there exits a possibility to extract of the SUSY
parameters from observable characteristics of the bursts and to
find constraints on the SUSY model. In the framework of Split
Higgsino scenario \cite{we} the neutralino mass $M_{\chi}\simeq
3\; \mathrm{TeV}$ is estimated from the statement that the Dark
Matter in the Universe has formed at the high symmetry phase of
the cosmological plasma. Neutralino of such type in the halo
annihilate into W- and Z-bosons and fermions (see
Fig.~\ref{fig:fig2}).
Total kinetic cross section is
\begin{eqnarray}
 \displaystyle (\si v)_{ann}=
 \frac{g_2^4\,(21-40\cos^2{\theta_W}+34\cos^4{\theta_W})}{256\; \pi
 M_{\chi}^2\cos^4{\theta_W}}.
 \label{annhalo}
\end{eqnarray}
The clump annihilation time in this model is $t_{ann}\simeq 30
\;s.$

\section{Calculation of characteristic photon energies and
annihilation spectrum}

The energy spectrum $dN_{\gamma}(E_{\gamma})/dE_{\gamma}$ is
determined by the distribution in multiplicity of the secondary
hadrons resulted from the neutralino annihilation. We use of the
fact that the total secondary hadrons multiplicity is nearly twice
larger than the charged hadrons multiplicity \cite{Eidelman}. The
average multiplicity of secondary charged hadrons $\langle \tilde
n_{ch}\rangle(\sqrt{s})$ was studied in $e^+e^-, \, pp, \, p \bar
p, \, e^{\pm}p$ processes \cite{hera}. It was established that
this quantity is some universal function of energy
\[
 \displaystyle \tilde n_{ch}(\sqrt{s})=A+B\ln \sqrt s+C\ln^2\sqrt s,\quad\tilde
 n_{ch}\equiv \langle n_{ch}\rangle (\sqrt{s}/q_0)-n_0
\]
with experimentally fixed parameters
\[
 A=3.11\pm 0.08,\quad
 B=-0.49\pm 0.09,\quad C=0.98\pm 0.02.
\]
To choice a specific channel it is necessary to fix parameters
$q_0,\,n_0$. For neutralino annihilation $\displaystyle
q_{0(\chi\chi)}=1,\; n_{0(\chi\chi)}=0.$

Further it is supposed that the part of charged hadrons $\kappa
\equiv \langle n_{ch}\rangle /\langle n_{h} \rangle$ doesn't
depend on energy and has the value $\kappa \simeq 0.49$, which is
extracted from $Z$-peak data \cite{Eidelman}.

For characteristic photon energies generation the following
processes are most important $\pi^0 \to 2\ga ,\;\eta^0 \to 3\ga
,\;\eta^0\to 3\pi^0 \to 6\ga$. In the fermion annihilation channel
(Fig.~\ref{fig:fig2}) the total hadron multiplicity is described
by following logarithmic function with good accuracy:
\begin{eqnarray}
 \displaystyle \langle n^{ff}_h
 \rangle=\kappa^{-1}(A+B\ln 2M_\chi+C\ln^22M_\chi)\simeq
 149\quad\mathrm{for}\quad M_\chi\simeq 3 \;\mathrm{TeV}.
 \label{nff}
\end{eqnarray}
An average energy of neutral pions in neutralino annihilation
secondaries is $\bar E_{\pi^0}\simeq \bar E_{\eta^0}\simeq
2M_{\chi}/\langle n^{ff}_h \rangle$. Then in the $\pi^0\to 2\ga$
decay maximal characteristic energy of photon is equal:
\begin{eqnarray}
 \bar E_{\ga
 (\pi^0\to\, 2\ga)}\simeq \bar E_{\pi^0}/2=20 \;\mathrm{GeV}.
 \label{xe1}
\end{eqnarray}

Analogously, for reactions $\eta^0\to 3\ga$ and $\eta^0\to
3\pi^0\to 6\ga$ energies of photons are:
\begin{eqnarray}
 \bar E_{\ga (\eta^0\to\, 3\ga)}\simeq \bar E_{\eta^0}/3=13.5 \;\mathrm{GeV}\,,\qquad
 \bar E_{\ga (\eta^0\to\, 6\ga)}\simeq \bar E_{\eta^0}/6=6.7 \;\mathrm{GeV}.
 \label{xe2}
\end{eqnarray}

In the boson annihilation channel (Fig.~\ref{fig:fig2}) the total
hadron multiplicity of $W$- and $Z$-bosons decays is:
\begin{eqnarray}
 \displaystyle \langle n^{WZ}_h
 \rangle\simeq 42.9.
 \label{nwz}
\end{eqnarray}
Here neutral pion average energy is $\bar E_{\pi^0}\simeq \bar
E_{\eta^0}\simeq M_{\chi}/\langle n^{WZ}_h \rangle$ and maximal
characteristic photon energies are:
\begin{eqnarray}
  \begin{array}{c} \vspace*{5mm}
  \bar E_{\ga (\pi^0\to\, 2\ga)}\simeq \bar E_{\pi^0}/2\simeq 35
  \;\mathrm{GeV},
\\
  \bar E_{\ga (\eta^0\to\, 3\ga)}\simeq \bar E_{\eta^0}/3=23.3
  \;\mathrm{GeV}\,,\quad \bar E_{\ga (\eta^0\to\, 6\ga)}\simeq \bar
  E_{\eta^0}/6=12 \;\mathrm{GeV}.
  \end{array}
\label{xe3}
\end{eqnarray}

In a wide energy region the multiplicity distribution can be
described with a good accuracy by the Negative Binomial
Distribution (NBD) that depends on energy very weakly
(logarithmically) \cite{Eidelman}:
\begin{eqnarray}
 \begin{array}{c} \vspace*{5mm}
 \displaystyle P(n; \tilde n, k)=\frac{k(k+1)...(k+n-1)}{n!}\cdot
 \frac{(\tilde n/k)^n}{[1+(\tilde n/k)]^{n+k}}\;,
\\
 \displaystyle k^{-1}(\sqrt{s})= a+b\ln \sqrt s,
 \end{array}
 \label{nbd}
\end{eqnarray}
where $n\equiv n_{ch},\,\tilde n\equiv \tilde n_{ch}$. For various
channels the coefficients in the function $k^{-1}(\sqrt{s})$ are
different:
\begin{eqnarray}
 \displaystyle a_{e^+e^-}= -0.064\pm
 0.003,\qquad b_{e^+e^-}=0.023\pm 0.002;
 \label{abee}
\end{eqnarray}
\begin{eqnarray}
 \displaystyle
 a_{pp/\bar pp}= -0.104\pm 0.004,\qquad b_{pp/\bar pp}=0.058\pm
 0.001.
 \label{abpp}
\end{eqnarray}

\begin{figure}[!h]
 \centerline{\includegraphics[width=0.5\textwidth]{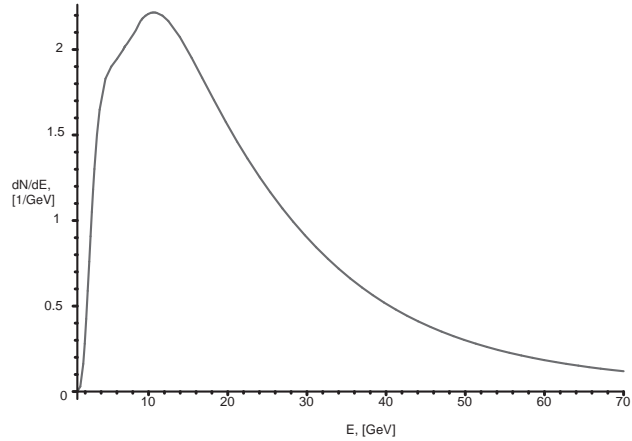}}
   \caption{\label{fig:fig3}
   \small Gamma ray annihilation spectrum.}
\end{figure}

Using the NBD it is possible to find the approximate energy
distribution of photons. For example, in the hadronic channel with
$Br (h)\simeq 0.58$ it is supposed that the branching ratios for
various hadrons $Br (i/h)\equiv \langle n_i\rangle/\langle
n_h\rangle$ are nearly constants and equal to the corresponding
branching ratios extracted from $e^+e^-$-annihilation
\cite{Eidelman}. Each annihilating neutralino pair gives a number
of neutral pions $Br(\pi^0/h)\cdot 2\bar n P(n; \bar n, k)$ with
the probability $Br (h)$. These pions generate the following
distribution of photons $\Delta n_\ga\simeq 2\cdot
Br(h)Br(\pi^0/h)\cdot 2\bar n P(n; \tilde n, k)\Delta n$ with
energy $E_\ga=M_\chi/2n$ in the multiplicity interval $\Delta n$,
which is connected with energy interval $\Delta n=M_\chi \Delta
E_\ga/ 2E_\ga^2$. For $\eta^0$-mesons and boson annihilation
channel considerations are analogous. Then the number of photons
with energy $E_{\gamma}$ per one annihilation act is:
\begin{eqnarray}
 \begin{array}{c} \vspace*{5mm}
 \displaystyle \frac{dN_\ga}{dE_\ga}\simeq
 \frac{2M_\chi}{E_\ga^2}\Large\left\{\displaystyle \left[ Br(\pi^0/h)+
 Br(\eta^0/h)\cdot Br(\eta^0\to 2\ga)\right]\times \right.
\\ \vspace*{5mm}
 \displaystyle \times \left[Br(h)\langle n^{ff}_{ch}\rangle
 P\left(\frac{M_\chi}{2E_\ga}; \langle n^{ff}_{ch}\rangle,
 k_{ff}\right)+\frac12 Br(WZ)\langle n^{WZ}_{ch}\rangle
 P\left(\frac{M_\chi}{4E_\ga}; \langle n^{WZ}_{ch}\rangle,
 k_{WZ}\right)\right]+
\\ \vspace*{5mm}
 \displaystyle+ Br(\eta^0/h)\left[Br(\eta^0\to 3\pi^0)+\frac13 Br(\eta^0\to
 \pi^+\pi^-\pi^0)\right]\times
\\ \vspace*{5mm}
 \times \displaystyle \left[Br(h)\langle n^{ff}_{ch}\rangle
 P\left(\frac{M_\chi}{6E_\ga}; \langle n^{ff}_{ch}\rangle,
 k_{ff}\right)+\frac12 Br(WZ)\langle n^{WZ}_{ch}\rangle
 P\left(\frac{M_\chi}{12E_\ga}; \langle n^{WZ}_{ch}\rangle,
 k_{WZ}\right)\right]+
\\ \vspace*{5mm}
 \displaystyle +\frac13 Br(\eta^0/h) Br(\eta^0\to \pi^+\pi^-\ga)\times
\\
 \displaystyle \left.\times \left[Br(h)\langle n^{ff}_{ch}\rangle
 P\left(\frac{M_\chi}{3E_\ga}; \langle n^{ff}_{ch}\rangle,
 k_{ff}\right)+\frac12 Br(WZ)\langle n^{WZ}_{ch}\rangle
 P\left(\frac{M_\chi}{6E_\ga}; \langle n^{WZ}_{ch}\rangle,
 k_{WZ}\right)\right]\right\}.
 \label{dN/dE}
 \end{array}
\end{eqnarray}
Here $Br(WZ)\simeq 0.2$ is the total branching ratio for
neutralino annihilation into $W$- and $Z$-bosons; charge hadron
multiplicities $\langle n^{ff}_{ch}\rangle$ and $\langle
n^{WZ}_{ch}\rangle$ are defined in (\ref{nff}) and (\ref{nwz});
parameters $k^{-1}_{ff}=k^{-1}(2M_{\chi})=0.4$ and
$k^{-1}_{WZ}=k^{-1}(M_{\chi})=0.12$ were determined with
coefficients from (\ref{abpp}) and (\ref{abee}), respectively. The
spectrum is shown in Fig.~\ref{fig:fig3}.

The particular duration of the GRB and the specific shape of its
energy spectrum are the characteristic features of considering
astrophysical event -- the collapse and total annihilation of the
neutralino clump. Calculated values of characteristic photon
energies are close to the threshold of sensitivity for the
satellite detector GLAST. Possibility of neutralino annihilation
spectrum registration at this apparatus will be clarified when the
GLAST will be put into operation.

\section{Conclusion}
We have proposed the new production mechanism of the cosmological
gamma-ray bursts. In the framework of this mechanism the gamma-ray
burst represents as a result of collapse and total annihilation of
the neutralino clump. The GRB signal when registered should
contain an information on SUSY parameters and the details of the
annihilation process. We have calculated the characteristic photon
energies and annihilation spectrum for $M_{\chi}=3\,\mathrm{TeV}$
in the framework of Split Higgsino scenario. The particular
duration of the GRBs and specific shape of its energy spectrum are
the characteristic features which can help to separate considering
astrophysical events from the radiation background. From
astrophysical data on GRBs with high-energy photons it is possible
to extract an information about SUSY parameters in different
models. It can help to investigate compatibility of the SUSY model
with cosmology, cosmogony and high-energy collider data.

\end{document}